\documentstyle[twocolumn,prl,aps]{revtex}
\begin{document} 
\addtolength{\textheight}{8mm}
\title{
Renormalization of the P- and T-odd nuclear potentials\\
by the strong interaction and enhancement \\ of P-odd effective field
}
\author{V. V. Flambaum and O. K. Vorov}
\address{$^*$ Theoretical Physics Department, School of Physics,
University of New South Wales, Sydney,
NSW 2052, Australia
\\
(Received  7 September 1993)
}
\address{\mbox{ }}
\address{\mbox{ }}
\address{\parbox{14cm}{\rm \mbox{ }\mbox{ }
Approximate analytical formulas for the self-consistent 
renormalization
of P,T-odd and P-odd weak nuclear potentials by the residual 
nucleon-nucleon strong interaction are derived. The contact spin-flip 
nucleon-nucleon
interaction reduces the constant of the P,T-odd potential 1.5 times 
for the proton
and 1.8 times for the neutron. Renormalization of the P-odd potential
is caused by the velocity
dependent spin-flip component of the strong interaction. 
In the standard variant 
of $\pi + \rho$-exchange, the conventional strength values lead to 
anomalous
enhancement of the P-odd potential.
Moreover, the $\pi$-meson exchange contribution seems 
to be large enough to generate 
an instability (pole) in the nuclear response to a weak potential.
}}
\address{\mbox{ }}
\address{\parbox{14cm}{\rm PACS numbers: 21.30.+y, 13.75.Cs, 24.80.Dc
}}
\maketitle
\makeatletter
\global\@specialpagefalse
\def\@oddhead{REV\TeX{} 3.0
\hspace{1.0cm}
PHYSICAL REVIEW  C \hspace{1.5cm} VOLUME 49, NUMBER 4 , p.1827 
\hspace{2.0cm} APRIL 1994}
\let\@evenhead\@oddhead
\makeatother
Recent measurements of the effects of parity nonconservation (PNC)
in nuclear reactions produced several results which still have 
not been
explained: permanent sign PNC effects in neutron capture by 
$^{232}Th$
\cite{FRANKLE} and very large PNC effects in Moessbauer 
transitions \cite{Tsin}.
One can consider these observations as a hint that there could 
be some 
new mechanisms
to enhance the weak interaction in the nucleus. Therefore,
it is time to consider possible corrections which can 
influence the
magnitude of PNC effects. In our work \cite{FV} it was 
pointed out that
the  residual strong interaction can enhance the two-nucleon 
PNC-interaction 
$\sim A^{1/3}$ times ($A$ is a nucleon number). We
called the residual interaction which combines the action of 
the weak potential
with the residual strong interaction the 
induced parity nonconserving interaction
(IPNCI). However, the dominating part of the two-nucleon IPNCI, 
which was produced 
by the velocity-independent contact strong interaction, 
does not contribute to the
single-particle weak potential. In the present work we consider 
the part of the
strong interaction which renormalizes the single-particle 
weak potential.
Renormalization of the P- and T-odd nuclear potential which 
contributes 
to P- and
T-odd nuclear moments and to P- and T-odd effects in neutron 
scattering 
is also considered.

Let us start from the consideration of P- and T-odd nuclear 
potentials 
(see , e.g. Ref.\cite{FKSNP}):
\begin{eqnarray}
\label{p1}
H_{TP}=\frac{G}{2 \sqrt{2} m} \eta ({\bf \sigma} {\bf \nabla}) 
\rho \simeq \theta 
{\bf \sigma}{\bf \nabla} U, 
\nonumber\\
\quad
\theta = \eta \frac{G}{2 \sqrt{2} m} \frac{\rho(0)}{U(0)}=
-2 \cdot 10^{-8} 
\eta \cdot fm,
\end{eqnarray}
where ${\bf \sigma}$ is the doubled nucleon spin, $\rho$ is 
nuclear density,
$m$ is proton mass, $G$ is the Fermi constant, $\eta$ is a 
dimensionless constant
characterizing the strength of T- and P-odd interaction (
the limits on these constants for protons ($\eta_{p}$) 
and neutrons ($\eta_{n}$) 
were obtained from atomic \cite{ATOM} and molecular \cite{MOLECULE}
electric dipole
moment measurements), and $U$ is the strong nuclear potential, 
$fm=10^{-13}cm$.
The shape of the potential $U$ and the nuclear density $\rho$ 
is known to be 
approximately similar. We used this fact in Eq.(\ref{p1}). 
Correspondingly, the whole
potential affecting the nuclear motion is equal to
\begin{equation}
\label{p2}
\bar U = U + H_{TP} = U({\bf r}) + 
\theta {\bf \sigma \nabla} U \simeq U({
\bf r} + \theta {\bf \sigma}). 
\end{equation}
Hence, it is obvious that the nucleon wave function with 
the $H_{TP}$ 
taken into
account has the form:
\begin{equation}
\label{p3}
\psi = \psi({\bf r}+\theta{\bf \sigma}) = 
(1+ \theta {\bf \sigma \nabla})
\psi({\bf r}) = \psi + \delta \psi,
\end{equation}
where $\psi({\bf r})$ is the nonperturbed wave function.
The direct correction to the strong potential induced by 
a small perturbation
can be written as follows:
\begin{eqnarray}
\label{p4}
\delta V(1) = \sum_{a} \int d2 
[\delta \psi^{+}_{a}(2) V(1,2)
\psi_{a}(2)+
\nonumber\\
+ \psi^{+}_{a}(2) V(1,2) \delta \psi_{a}(2)].
\end{eqnarray}  
Here the notation $1(2) \equiv \lbrace {\bf r}_{1(2)}, 
\sigma_{1(2)}, \tau_{1(2)}
\rbrace$
stands for the full set of the  nucleon variables (coordinate, 
spin and
isospin) 
and the summation is carried out over the occupied nucleon states $a$.

We use the Landau-Migdal  parametrization of the strong interaction
\begin{equation}
\label{p5}
V({\bf r}_,
{\bf r'})=C\delta({\bf r}-{\bf r}')[f_{0}+f'_{0}{\bf \tau}
{\bf \tau'}+
g{\bf \sigma}{\bf \sigma'}+g'{\bf \tau}{\bf \tau'}{\bf \sigma}
{\bf \sigma'}],
\end{equation}
where $C=300$ $MeV \cdot fm^{3}$, $g=0.575, g'=0.725$ 
and only the direct terms are considered
(see e.g. Refs.\cite{Migdal},\cite{BET}). Using Eq.(\ref{p3}) for 
$\psi + \delta
\psi$  and integration by parts in Eq.(\ref{p4}), we otain the correction 
to the T,P-odd potential:
\begin{eqnarray}
\label{p6}
\tilde H_{TP}= - \sum_{a} \theta_{2} \int d^{3}r \psi^{+}_{a}
({\bf r}_{2})
\lbrack {\bf \sigma_{2} \nabla_{2}}, V({\bf r}_{1},{\bf r}_{2}) 
\rbrack
\psi_{a}({\bf r}_{2}) = \nonumber\\
=\sum_{a} \theta_{2} (g+g'{\bf \tau}_{1}{\bf \tau}_{a}) 
({\bf \sigma}_{1}
{\bf \nabla}) |\psi_{a}|^{2} = \gamma {\bf \sigma \nabla} \rho , 
\end{eqnarray}
$\gamma= C (\theta_{p} \frac{Z}{A} (g \pm g')+\theta_{n} \frac{N}{A} 
(g \mp g'))$ for protons (neutrons).
Hereafter, $[ , ]$ means commutator,
and $\rho = \sum_{a} |\psi_{a}|^{2}$.
We put proton density $\rho_{p}=\frac{Z}{A} \rho$,
neutron density $\rho_{n}=\frac{N}{A} \rho$ and 
$<{\bf \sigma}_{2}>=0$ 
(we consider the potential created by paired nucleons). 
Now we should solve
the self-consistent equation $H_{TP}=H^{0}_{TP}+ \tilde H_{TP}$ 
for
T-,P-odd potential: 
\begin{equation}
\label{p7}
\theta \sigma \nabla U = \theta^{0} \sigma \nabla U + 
\gamma \frac{\rho(0)}{U(0)}   
\sigma \nabla U.
\end{equation}
Here $H^{0}_{TP}$ contains the ``initial'' values of the T-,P-odd 
interaction
constants $\eta^{0}_{p}$ and $\eta^{0}_{n}$ (or $\theta^{0}_{p}$ 
and
$\theta^{0}_{n}$), while $H_{TP}$ and $\tilde H_{TP}$ 
contain ``final'' values 
of the
constants. The solutions for the pair of  simple linear algebraic 
equations for     
the constants are the following:
\begin{eqnarray}
\eta_{p}=\frac{1}{D} (\eta^{0}_{p}[1+ \tilde C (g+g')N/A]-
\eta^{0}_{n} 
\tilde C (g-g') N/A ) \simeq \frac{\eta^{0}_{p}}{1.5}, 
\nonumber\\
\eta_{n}=\frac{1}{D} (\eta^{0}_{n}[1+ \tilde C (g+g')Z/A]-
\eta^{0}_{p} 
\tilde C (g-g') Z/A ) \simeq \frac{\eta^{0}_{n}}{1.8},  
\nonumber\\ 
D=[1+ \tilde C (g+g')N/A][1+ \tilde C (g+g') Z/A] - 
\qquad \qquad  
\nonumber\\
{\tilde C}^{2}(g-g')^{2}
Z N/A^{2}, 
\qquad \qquad \qquad  
\nonumber
\end{eqnarray}
Here, ${\tilde C} = C \rho / |U| = \frac{4}{3} \frac
{\varepsilon_{F}}{|U|}=
\frac{4}{3}(1+\frac{|\varepsilon|}{\varepsilon_{F}})^{-1} 
\simeq 1$ and
$\eta^{0}_{p}$ and $\eta^{0}_{n}$ are  
the initial values of the constants. We used the well known 
relations :
\begin{equation}
\label{p8}
C=\frac{\pi^{2}}{p_{F} m}, \quad \rho= \frac{2p_{F}^{3}}
{3\pi^{2}}, \quad
\varepsilon_{F}=\frac{p_{F}^{2}}{2m}, \quad |U|= \varepsilon_{F}+
|\varepsilon|, 
\end{equation}
where $p_{F}$ is a Fermi momentum, $|\varepsilon|$ is a nucleon 
separation
energy. We also have taken into account in the numerical 
estimate that $|g-g'|$ is small. 

Thus, the strong residual interaction reduces the values of the 
T-,P-odd
potential constants $1.5 \div 1.8$ times. Note that the 
response of the 
nucleus to the T- and P-odd potential (\ref{p1}) as a function of 
the interaction constants
has poles ($D=0$) at $g={\tilde C}^{-1} \simeq -1$ and $g' \simeq
{\tilde C}^{-1} \simeq -1$ (for $N \simeq Z$). 
The positions of the poles differ from the instability points 
in an infinite 
Fermi system $g = g'= -1.5$ (see, e.g., Refs. \cite{K-S},
\cite{P-N})
since the 
interaction (\ref{p1}) does not exist in the infinite system ($H_{PT}=0$ 
at $\rho=const$). 

It is interesting that the T- and  P-odd interaction induces a 
spin hedgehog 
(${\bf \sigma} \sim {\bf r}$) in the nucleon spin distribution 
within a 
spherical
nucleus. A simple calculation with the wave function (\ref{p3}) gives 
the following
proton and neutron spin distributions:
\begin{equation}
\label{p9}
\sigma_{p}({\bf r}) = \theta_{p} \nabla \rho_{p}({\bf r}), \quad
\sigma_{n}({\bf r}) = \theta_{n} \nabla \rho_{n}({\bf r}) 
\end{equation}      
The interaction $\tilde H_{TP}$ in Eq.(\ref{p6}) is, in fact, a 
strong interaction of the 
nucleon with the spin hedgehog ($C g {\bf \sigma \sigma(r)}$).

Now we turn to considering corrections to the weak P-odd and 
T-even potential
\begin{equation}
\label{p10}
W=\frac{G}{2\sqrt{2}m} g^{W} ( {\bf \sigma p} \rho + 
\rho{\bf \sigma p}).
\end{equation}
Here $p$ is the nucleon momentum, the dimensionless constants 
$g^{W}_{p}$ for the proton and 
$g^{W}_{n}$ (for the neutron) are of order of unity (the 
notation $\varepsilon \simeq 
1.0 \cdot 10^{-8} g$ is also adopted in the current literature). 
In a simple model 
of a constant nuclear density it is easy to find the result of 
the action
of the perturbation $W$ (see Ref. \cite{MICHEL}, and the first
paper of Ref.\cite{FKSPL}):
\begin{eqnarray}
\label{p11}
\tilde \psi = exp(-i \xi {\bf \sigma r}) \psi({\bf r}) \simeq
(1-i \xi {\bf \sigma r}) \psi({\bf r})  ,
\nonumber\\
\quad \xi= \frac{G}{\sqrt{2}} g^{W} \rho = \varepsilon m. \qquad \qquad
\end{eqnarray}
In the general case (real density shape and spin-orbit 
interaction taken into
account) the correction to the wave function contains an extra 
spherically 
symmetric function $\varphi_{a}(r)$ (see, e.g. Ref. \cite{KH}):
\begin{equation}
\label{p12}
\delta \tilde \psi_{a} = - i ({\bf \sigma r}) \varphi_{a}(r)
\psi_{a}({\bf r}).
\end{equation}
The P-odd weak interaction (\ref{p10}) also changes the 
spin distribution. It rotates the spin
around vector ${\bf r}$ [see Eq.(\ref{p11})] by the angle $\xi r$ and 
creates
a spin spiral
\cite{KH}. However, after the summation over paired nucleons 
this spin structure
disappears.
As a result, the contact spin-dependent strong interaction
(\ref{p5}) does not contribute to the renormalization of the weak potential 
(because of the factor $i$ in Eqs.(\ref{p11}) and (\ref{p12}) 
the contribution of 
($\delta \psi^{+}$)
compensates the contribution from ($\delta \psi$) in Eq.(\ref{p4}) for 
the correction
to the potential). This result looks natural since the only possible 
orientation of the spin in the spherical nucleus ${\bf \sigma} 
\sim {\bf r}$ 
violates both P- and  T-invariances and can not be produced by a 
T-even weak 
interaction (\ref{p10}). 

The correlation which is actually produced by the weak interaction 
is
${\bf \sigma p}$. To reveal such structures the strong interaction 
must
be spin and momentum dependent as well 
(another possibility is related to a finite range exchange
interaction; it will be considered below). Within Landau-Migdal 
theory, the 
momentum
dependence is usually described \cite{K-S} by the 
following extra term
in (\ref{p5}):
\begin{eqnarray}
\label{p13}
V_{1}= \frac{1}{4} C p_{F}^{-2} (g_{1}+g'_{1} {\bf \tau}_{1}
{\bf \tau}_{2})
({\bf \sigma}_{1}{\bf \sigma}_{2})
\nonumber\\
\biggl[ {\bf p}_{1} {\bf p}_{2} \delta({\bf r}_{1}-{\bf r}_{2}) +
{\bf p}_{1} \delta({\bf r}_{1}-{\bf r}_{2}) {\bf p}_{2} + 
\nonumber\\
{\bf p}_{2} \delta({\bf r}_{1}-{\bf r}_{2}) {\bf p}_{1} +
\delta({\bf r}_{1}-{\bf r}_{2}){\bf p}_{1} {\bf p}_{2} \biggr]. 
\end{eqnarray}
The constants of this interaction are found 
to be $g_{1}=-0.5$, $g'_{1}=-0.26$ (Ref.\cite{K-S}). 
Using Eqs.(\ref{p4}),
(\ref{p11}) and
(\ref{p13}) we can calculate the corresponding correction to 
the weak potential:
\begin{eqnarray}
\label{p14}
\tilde W = \sum_{a} \int d^{3} r_{2} \psi^{+}_{a}({\bf r}_{2}) 
\lbrack i \xi_{a} {\bf \sigma}_{2} {\bf r}_{2},V_{1} \rbrack 
\psi_{a}
({\bf r}_{2}) =  \nonumber\\
=-\sum_{a} \xi_{a}(g_{1}+g'_{1} {\bf \tau}_{1}{\bf \tau}_{a})
\frac{C}{2p_{F}^{2}} \lbrace {\bf \sigma}_{1} {\bf p}_{1}, 
|\psi_{a}|^{2}
\rbrace = 
\nonumber\\ =
Q [ ({\bf \sigma p}) \rho + \rho ({\bf \sigma p})] , 
\end{eqnarray}
where
$Q= - \frac{C}{2p_{F}^{2}} \lbrack \frac{Z}{A} (g_{1} 
\pm g'_{1})\xi_{p}+
\frac{N}{A} (g_{1} \mp g'_{1})\xi_{n} \rbrack$
for protons (neutrons) correspondingly.
A self-consistent solution of the equation for the total P-odd 
nuclear potential 
$W=W^{0}+ \tilde W$ gives the following values of the potential 
constants:
\begin{eqnarray}
\label{p15}
g^{W}_{p}=\frac{1}{D} \lbrace
g^{W0}_{p} [1+ \frac{2N}{3A}(g_{1}+g'_{1})] - 
\frac{2N}{3A}g^{W0}_{n}(g_{1}-g'_{1}) \rbrace , \nonumber\\
g^{W}_{n}=\frac{1}{D} \lbrace
g^{W0}_{n} [1+ \frac{2Z}{3A}(g_{1}+g'_{1})] - 
\frac{2Z}{3A}g^{W0}_{p}(g_{1}-g'_{1}) \rbrace,   \nonumber\\
D=[1+\frac{2N}{3A}(g_{1}+g'_{1})][1+\frac{2Z}{3A}(g_{1}+g'_{1})]
\nonumber\\
-
\frac{4NZ}{9A^{2}}(g_{1}-g'_{1})^{2}.
\end{eqnarray}
We have taken into account here that $C\rho m p_{F}^{-2}=2/3$. 
It is interesting
that the poles ($D=0$) in the response of a nucleus to the weak 
potential 
$W \sim {\bf \sigma p}$ coincide with the boundary of stability 
for a Fermi-
liquid with interaction (\ref{p13}): $g_{1}=g'_{1}=-1.5$ at $N=Z$ 
(see, e.g. Refs.
\cite{K-S},\cite{P-N}). This is not too surprising 
since we used
the approximation
$\rho = const$ to obtain the wave function (\ref{p11}). \footnote{     
We had known from a private communication with V.G.Zelevinsky
that he independently obtained a similar result: The 
correction to the effective
field ${\bf \sigma p}$ diverges at the same point where the 
first harmonic
of the Landau interaction $g_{1}({\bf \sigma}_{1}
{\bf \sigma}_{2})({\bf p}_{1}
{\bf p}_{2})$ leads to the instability of the Fermi liquid.}

The interaction $V_{1}$ with the constants $g_{1}=-0.5$, 
$g'_{1}=-0.26$
does not cause instability. However, it acts in the direction 
of the poles
and increases the P-odd potential:
\begin{equation}
\label{p16}
g_{p}=1.3 g^{0}_{p}+ 0.18g^{0}_{n}, \quad g_{n}= 1.4g^{0}_{n}+
0.12 g^{0}_{p}.
\end{equation}
Therefore, the Landau-Migdal interaction $V+V_{1}$ 
[Eqs.(\ref{p5}) and (\ref{p13})] 
does not 
produce crucial changes in the values of interaction constants 
for 
P,T-odd and P-odd potentials. The corrections are of the same 
size as, say, 
corrections to the Schmidt values of the magnetic moments.
In fact, the Landau-Migdal interaction originates from the 
underlying 
$\pi+\rho$-exchange interaction \cite{BSG} which generates 
also
tensor components. To account for the latter 
destroys this
``idyllic'' picture at least for the P-odd potential.
The rest of the paper is devoted to the calculation of the 
$\pi+\rho$ contribution.
         
In the present random-phase-approximation like calculations, 
the correction to the nucleon P-odd
potential $\tilde W = \sum_{\mu \nu} \delta w_{\mu \nu} 
\mu^{+}\nu$
due to the strong interaction
$\hat V =\frac{1}{2}\sum_{abcd}a^{+}c^{+}V_{abcd} d b$ is 
given by the 
expression
\begin{eqnarray}
\label{p17}
\delta w_{\mu \nu} = \sum_{ab}(A_{ab}V_{ba \mu \nu} -
V_{ab \mu \nu} A_{ba})n_{a}-
\nonumber\\
\sum_{ab}(A_{ab}V_{b \nu \mu a}-V_{a \nu \mu b}A_{ba})n_{a},   
\end{eqnarray}
where the first sum is the direct contribution and the second 
is exchange one
($\mu^{+},\mu, a^{+},a$ ... are 
creators and destructors of nucleons in the corresponding 
single-particle states), $n_{a}$ are the occupation numbers,
$n_{a}\equiv <a^{+}a>$,  
and $A_{ab}=<\psi_{a}|i\xi ({\bf \sigma r}) |\psi_{b}>$
are single-
particle matrix elements of the P-odd mixing operator [see Eq.(\ref{p11})]. 
For the $\pi + \rho$-
interaction \cite{BSG}, $\hat V_{abcd}$ is given by 
\begin{equation}
\label{p18}
V_{abcd}= \int d1 d2 \psi^{+}_{a}(1) \psi^{+}_{c}(2)V^{\pi+\rho}(1,2)
\psi_{b}(1) \psi_{d}(2),                        
\end{equation}
where $V^{\pi+\rho}(1,2)$, in $p$-representation, is
\begin{eqnarray}
\label{p19}
\qquad 
V^{\pi+\rho}(1,2)= \qquad \qquad
\nonumber\\
- 4 \pi ({\bf \tau_{1} \tau_{2}})
\left[ \frac{f_{\pi}^{2}}{ m_{\pi}^{2}}
\frac{({\bf \sigma_{1} q})({\bf \sigma_{2} q})}{q^{2}+m_{\pi}^{2}}+
\frac{f_{\rho}^{2}}{m_{\rho}^{2}}
\frac{[{\bf \sigma_{1} \times q}][{\bf \sigma_{2} \times q}]}
{q^{2}+m_{\rho}^{2}} \right],
\label{2.3}
\end{eqnarray}
where ${\bf q}$ is the momentum transfer,
$m_{\pi(\rho)}$ is the pion (rho-meson) mass, $f_{\pi}^{2}=
0.08$ is a pion coupling constant, and $f_{\rho}^{2}$ is
corresponding $\rho$-meson coupling ranging from 1.86 to 4.86 
(``weak'' and
``strong'' couplings correspondingly \cite{BSG}).
In the coordinate representation, the last expression becomes 
a potential
depending on $|{\bf r}_{1}-{\bf r}_{2}|$. Thus its commutator 
with 
$A=i \xi ({\bf \sigma r})$ in (\ref{p17}) (direct terms) is zero; 
while the
exchange terms contain (due to nonlocality of the potential) 
an effective 
velocity dependence and yield a nonzero contribution to $\tilde W$. 
To calculate 
the latter, we should reduce the exchange terms in (\ref{p17}) to a
direct form which require the change 
${\bf q} 
\rightarrow {\bf p}_{1}-{\bf p}_{2}$ 
(the nucleons are on the Fermi surface) and 
Fierz transformation of the spin and isospin tensor structures 
\cite{BET}. 
After performing that, we obtain, for $V_{b \nu \mu a}$,
\begin{displaymath}
V_{b \nu \mu a} = 
\int d1 d2 \psi^{+}_{b}(2) \psi^{+}_{\mu}(1)V'(1,2)\psi_{\nu}(1)
\psi_{a}(2),
\end{displaymath}
with $V'(1,2)$ being equal to
\begin{eqnarray}
\label{p20}
V'(1,2)=
\nonumber\\
-2 \pi \left( \frac{3}{2}-\frac{1}{2}
({\bf \tau_{1} \tau_{2}}) \right)
\sum_{\alpha \beta} [2 \sigma_{1 \alpha} \sigma_{2 \beta} 
+(1-({\bf \sigma_{1} \sigma_{2}})) \delta_{\alpha \beta} \rbrack
\nonumber\\
\times
\biggl[ \frac{f_{\pi}^{2}}{m_{\pi}^{2}}
\frac{({\bf p_{1}-p_{2}})_{\alpha}({\bf p_{1}-p_{2}})_{\beta}}
{({\bf p_{1}-p_{2}})^{2}+m_{\pi}^{2}}+
\nonumber\\
\frac{f_{\rho}^{2}}{m_{\rho}^{2}}
\frac{\delta_{\alpha \beta}({\bf p_{1}-p_{2}})^{2} -
({\bf p_{1}-p_{2}})_{\alpha}({\bf p_{1}-p_{2}})_{\beta}}
{({\bf p_{1}-p_{2}})^{2}+m_{\rho}^{2}}
\biggr]. \label{2.4}
\end{eqnarray}
By use of that, the second sum in Eq.(\ref{p17}) is reduced to the 
expectation
value of the commutator 
$[i \xi_{2}({\bf \sigma}_{2}{\bf r}_{2}) ,V'(1,2)]$ 
and we obtain the 
meson exchange correction
$\tilde W$ to the P-odd potential acting on the first nucleon: 
\begin{equation}
\label{p21}
\tilde W^{\pi+\rho} =
 - <\psi^{+}_{a}(2)|[i \xi({\bf \sigma}_{2} {\bf r}_{2}),V'(1,2)]
|\psi_{a}(2)>.
\end{equation}
Here notation $<...>$ stands for the expectation value taken in the 
subspace of the wave functions of the core nucleons (label 2),
and the
summation is assumed over the states $a$ occupied by the 
nucleon 2. 
Calculating the commutator in (\ref{p21}) using (\ref{p20}) and the relations 
$[{\bf r}_{2 \alpha},{\bf p}_{2 \beta}]=i \delta_{\alpha \beta}$, 
$\langle \sigma_{2 \alpha}\sigma_{2 \beta} \rangle = \delta_{\alpha \beta}$, 
we obtain
\begin{eqnarray}
\label{p22}
\tilde W^{\pi+\rho} = \qquad \qquad \qquad \qquad \qquad  
\nonumber\\ =
K_{\pi} 
<\psi^{+}_{a}(2)| ({\bf \sigma}_{1}({\bf p}_{1}-{\bf p}_{2}))
\biggl[ \frac{1}
{({\bf p}_{1}-{\bf p}_{2})^{2}+m_{\pi}^{2}}-
\nonumber\\
\frac{1}{3} \frac{({\bf p}_{1}-{\bf p}_{2})^{2}}
{[({\bf p}_{1}-{\bf p}_{2})^{2}+m_{\pi}^{2}]^{2}} \biggr] |\psi_{a}(2) 
> \nonumber\\
-K_{\rho} 
<\psi^{+}_{a}(2)| ({\bf \sigma}_{1}({\bf p}_{1}-{\bf p}_{2}))
\nonumber\\
\left[ \frac{1}
{({\bf p}_{1}-{\bf p}_{2})^{2}+m_{\rho}^{2}}-
\frac{1}{2} \frac{({\bf p}_{1}-{\bf p}_{2})^{2}}
{[({\bf p}_{1}-{\bf p}_{2})^{2}+m_{\rho}^{2}]^{2}} \right] 
|\psi_{a}(2)>,
\end{eqnarray}
where the right hand side remains to be an operator acting on the wave
functions of nucleon 1, and the constants are 
$K_{\pi}= 6 \pi (f_{\pi}^{2}/m_{\pi}^{2})
(3-({\bf \tau_{1} \tau_{2}})) \xi_{2}$, 
$K_{\rho}= 8 \pi (f_{\rho}^{2}/m_{\rho}^{2})
(3-({\bf \tau_{1} \tau_{2}})) \xi_{2}$.
To evaluate the expression (\ref{p22}), one 
can employ, e.g., the Fermi-gas approximation to parametrize the 
density of core 
nucleons $\sum_{a} \psi^{+}_{a}(2)\psi_{a}(2)$ as has been widely used 
in such
calculations [e.g., in obtaining the ``bare'' nucleon 
P-odd potential
\cite{AH}]. 
We obtain from Eq.(\ref{p22})
\begin{equation}
\label{p23}
\tilde W^{\pi+\rho} = 2Q^{\pi+\rho} \rho_{0} ({\bf \sigma_{1}p_{1}}), 
\label{2.7}
\end{equation}
where the constant $Q^{\pi+\rho}$ for proton and neutron has the 
following form:
\begin{eqnarray}
\label{p24}
Q^{\pi+\rho}_{p}=q(\xi_{p}\frac{Z}{A}+2\xi_{n}\frac{N}{A}),  \quad
Q^{\pi+\rho}_{n}= q(\xi_{n}\frac{N}{A}+2\xi_{p}\frac{Z}{A}), 
\nonumber\\
q=6 \pi (\frac{f_{\pi}^{2}}
{m_{\pi}^{4}}
W_{\pi}(\frac{p_{F}}{m_{\pi}})-\frac{4}{3}\frac{f_{\rho}^{2}}
{m_{\rho}^{4}}
W_{\rho}(\frac{p_{F}}{m_{\rho}}))
\end{eqnarray}
and  the nonlocality factors $W$ 
($W_{\pi,\rho} \rightarrow 1$ for 
$m_{\pi,\rho} \rightarrow \infty$)
are $W_{\pi}(\frac{p_{F}}{m_{\pi}})=0.11$,
$W_{\rho}(\frac{p_{F}}{m_{\rho}})=0.69$ for pion and $\rho$-meson
correspondingly.
The nonlocality effect is greater for the pion due to its
smaller mass ($m_{\pi}=0.7 fm^{-1}$ compared to 
$p_{F} \simeq 1.3 fm^{-1}$,
while $m_{\rho}=3.7 fm^{-1}$).
The above value $W_{\pi}$ is quite close to the result 
$W_{\pi}=0.16$
for the nonlocality
factor for the ``bare'' weak potential obtained in   
$\alpha$-cluster 
calculations \cite{FKSPL}.

To obtain the renormalization of the P-odd weak potential 
with account
for $\pi+\rho$-exchange, one should
use $\tilde W^{\pi+\rho}$ instead of $\tilde W$ in Eq.(\ref{p15}) 
in the self-consistent determination of $W$. With account for
that, the renormalization equations of the potential constants 
$g_{p,n}$ (\ref{p15})
take the form
\begin{eqnarray}
\label{p25}
g^{W}_{p}=\frac{1}{D} \lbrace 
g^{W0}_{p} [1- \frac{N}{A} k] + 
2 \frac{N}{A}g^{W0}_{n} k \rbrace , \nonumber\\
g^{W}_{n}=\frac{1}{D} \lbrace 
g^{W0}_{n} [1- \frac{Z}{A} k] + 
2 \frac{Z}{A}g^{W0}_{p} k \rbrace,   
\end{eqnarray}
where $k=2 q \rho m$ and, 
in that case, the
determinant $D$ is equal to
\begin{equation}
\label{p26}
D=( 1-\frac{N}{A} k )( 1-\frac{Z}{A} k)-\frac{NZ}{A^{2}} 4 k^{2}.
\end{equation} 
It is seen from the last term in Eq.(\ref{p24}) that the contribution from
$\rho$-meson exchange tends to compensate the effect of the 
$\pi$-meson, 
whereas the latter strongly pushes the solution (Eq.(\ref{p25})) 
in the direction
of the pole ($D=0$).
The equation $D=0$ determines a curve 
(function of $N/A$) 
corresponding to the border of stability
of the nuclear response to the
P-odd field. 
For real nuclei ($N/A \simeq 0.5 \div 0.6$) the position of the pole
corresponds to the critical value of $k=k_{c}=0.67$. 
The $\pi$-meson alone (with no $\rho$-exchange)
gives $k=k_{\pi} \simeq 1$
and produces instability 
in the ``shell-model'' nucleus. The $\rho$-meson exchange reduces 
the value
of $k$: ``strong'' $\rho$-meson coupling ($f^{2}_{\rho}=4.86$) gives
$k=0.4$
which corresponds to enhancement factors $g^{W}_{p}/g^{W0}_{p}=1.6$, 
$g^{W}_{n}/g^{W0}_{p}=0.7$  
(for $g^{W0}_{p}=4$ and $g^{W0}_{n} \simeq 0$, see, e.g., 
\cite{FKSPL}). 
Thus $g^{W}_{n} \sim g^{W}_{p}$
even for very small initial values of $g^{W0}_{n}$.  
``Weak'' coupling ($0.4 f^{2}_{\rho}$) gives 
$k \simeq 0.7 \simeq k_{c}$
(``infinite'' enhancement).
Of course, the accuracy of the present consideration
is not sufficient to give a definite answer in this situation 
(note that we
have considered the linear response only and neglected fine 
effects like
smoothing of the pion in nuclear matter \cite{Migdal},\cite{FST}), 
besides 
the uncertainty in $\pi$ and $\rho$ coupling constants in a nucleus. 
At least one can
say that 
$D \simeq 0$ means a possibility of strongly enhanced P-odd effects.    

Interpretation of this fact, resulting
mostly from the strong $\pi$-meson exchange contribution,  is not
straightforward:
Definitely, it is related to the question of the stability of 
a nucleus
under the tensor $\pi$-exchange interaction which has been 
already widely
discussed  
in the literature \cite{BET}, in particular, in relation 
to the problem of 
$\pi$-condensation in nuclei (see e.g. \cite{Migdal}, \cite{FST}).
Also, a large enhancement factor is naturally associated with the 
low-lying $0^{-}$-excitation (a pole in $D$ for nonzero frequency
of the PNC field).
The influence of the $0^{-}$-resonance on the PNC-effects 
was discussed
in Refs. \cite{KMF}, \cite{A-B}.
On the other hand, some effects decreasing the role of the 
$\pi$-exchange interaction 
in a real nucleus may exist and the stability of the nucleus 
may be restored
(e.g., due to a particular shell structure). At least, our 
consideration
proves that a possible mechanism exists leading to the 
enhancement of the
nuclear P-odd weak potential which is caused by the velocity 
dependent spin-flip 
component of the conventional residual strong
interaction with the standard values of its constants.    
Thus, new reliable experimental information on the P-odd
nuclear effects
would be desirable. In  view of the present considerations, 
it might be 
important not only for the weak interaction theory, but also 
for the study of strong interaction effects and nuclear 
structure. 

To conclude, we have considered the renormalization of 
the nuclear T,P-odd
and P-odd potentials due to the residual strong interaction 
in the Landau-
Migdal parametrization and with account for the tensor component 
of the
one $\pi,\rho$-meson exchange. The 
T,P-odd potential is found to be renormalized
moderately, while the renormalization of the P-odd potential 
proves to be 
greater and the tensor velocity dependent interaction, with 
standard values of the parameters, turns out to 
be able to produce a substantial enhancement, ``driving'' 
the solution for 
the self-consistent P-odd field towards the region of instability.

\noindent
\end{document}